# Porosity-mediated High-performance n-type Mg$_3$Sb$_2$ Based Thermoelectric Materials

*Xiaoxi Chen, Siyi Chang, Jin Chen, Pengfei Nan, Xiyang Li, Xu Chen, Xiaoshan Zhu, Binghui Ge\*, Wei Cai, Jiehe Sui\*, Shuqi Zheng, Fangwei Wang, Xiaolong Chen, Huaizhou Zhao\*.*

X.X. Chen, S. Chang, J. Chen, P. Nan, X. Li, X. Chen, X. Zhu, Prof. B. Ge, Prof. F. Wang, Prof. X.L. Chen, Prof. H. Zhao
Beijing National Laboratory for Condensed Matter Physics, Institute of Physics, Chinese Academy of Sciences, Beijing 100190, China
E-mail: bhge@iphy.ac.cn (B.G.); hzhao@iphy.ac.cn (H.Z.)
X.X. Chen, Prof. W. Cai, Prof. J. Sui
State Key Laboratory of Advanced Welding and Joining, Harbin Institute of Technology, Harbin 150001, China
E-mail: suijiehe@hit.edu.cn
S. Chang, Prof. S. Zheng
State Key Laboratory of Heavy Oil Processing, Department of Materials Science and Engineering, China University of Petroleum, Beijing 102249, China

Abstract: Whether porosity can effectively improve thermoelectric performance is still an open question. Herein we report that thermoelectric performance can be enhanced by creating porosity in n-type Mg$_{3.225}$Mn$_{0.025}$Sb$_{1.5}$Bi$_{0.49}$Te$_{0.01}$, with a *ZT* of ~0.9 at 323 K and ~1.6 at 723 K, making the average *ZT* much higher for better performance. The large improvement at room temperature is significant considering that such a *ZT* value is comparable to the best *ZT* at this temperature in n-type Bi$_2$Te$_3$. The enhancement was mainly from the improved electrical mobility and multi-scale phonon scattering, particularly from the well-dispersed bismuth nano-precipitates in the porous structure. We further extend this approach to half-Heuslers Nb$_{0.56}$V$_{0.24}$Ti$_{0.2}$FeSb and Hf$_{0.25}$Zr$_{0.75}$NiSn$_{0.99}$Sb$_{0.01}$ showing similar improvements, further advancing thermoelectric materials for applications.

Main Text: Thermoelectric (TE) device aims to directly convert thermal energy into electricity or vice-versa, with their efficiency being primarily governed by the engineering (*ZT*)$_{\text{eng}}$[1] or average (*ZT*)$_{\text{ave}}$.[2] Here, *ZT* is defined as $ZT = [S^2\sigma/(\kappa_L + \kappa_e)]T$, where *S*, σ, $\kappa_L$, $\kappa_e$, and *T* are the Seebeck coefficient, electrical conductivity, lattice thermal conductivity, electrical thermal conductivity, and absolute temperature, respectively. Among them, *S* and σ are counter-related through charge carrier concentrations, whereas $\sigma$ and $\kappa_e$ are coupled through the Wiedemann-Franz relationship, and $\kappa_L$ subjects to structure manipulation, which makes any enhancement of ZT quite challenging. However, owing to the recent extensive research, both power factor ($PF = S^2\sigma$) enhancement and strong phonon scattering have been successfully observed in benchmark thermoelectric materials.[3,4] The all-scale hierarchical structure spanning from atomic defects and the nanoscale endotaxial precipitates to the mesoscale grain boundary has resulted in a great reduction of $\kappa_L$ in Sr and Na co-doped PbTe, leading to a high ZT value of ~2.2.[5] Even higher ZT values, yet to be reproduced, have recently been reported in SnSe due to bond anharmonicity limiting the phonon lifetime in the time domain instead of the space domain, favoring its intrinsically low $\kappa_L$, and a couple of record ZTs were reported in both p- and n-type doped SnSe.[6,7] Another paradigm of phonon engineering includes the so-called "phonon-liquid electron-crystal," as found in the family of copper chalcogenides.[8] In terms of the electrical transport properties, the resonant state and band convergence schemes aim at enhancing the degree of band degeneracy $N_v$ in the electronic band structures to boost the weighted mobility $\mu_w$, which is proportional to the dimensionless thermoelectric material quality factor $B = (\frac{k_B}{e})^2 \frac{e(2m_e k_B T)^{3/2}}{3\pi^2 \hbar^3} \frac{\mu_w}{k_L} T$. The high performance in n-type $Mg_3Sb_2$ based materials,[9] $Mg_2(Sn,Si)$,[8] and "pseudo-cubic" materials mainly originate from their high $\mu_w$.[8] Meanwhile, a recent report shows that spin can add another degree of freedom in tuning the electrical transport in the so-called phonon-glass electron-crystal skutterudite In-filled $CoSb_3$ through compositing with the superparamagnetic Fe, Co, and Ni nanophases.[10]

In spite of the significant progresses in enhancing the thermoelectric performance in the last decades, some vital challenges still remain, hindering the widespread application of TE materials. First, TE materials with high average ZTs are needed for practical applications regardless for power generation or refrigeration.[3] Second, the rare (Te) and toxic constituent elements (Pb) in most TE materials have been a significant concern in terms of both environmental and economic costs.

Porous structure can potentially serve as an ideal architecture for the decoupling of phonons and electrons to improve ZT.[11,12] Meanwhile, a few recent reports,[13–17] although preliminary, have unveiled the potential of a meso-porous structure in terms of cost reduction and TE performance enhancement. However, the porous structures in these materials are only case specific and unfortunately not tunable. Thus, whether porous TE materials are really useful is worth further exploration.

In this work, we proposed a general approach to fabricate bulk porous TE materials. The synthesis scheme is shown in Figure 1A, the details can be found in Supporting Information. Unlike the traditional spark plasma sintering (SPS) process, our specially designed graphite die allows the decoupling of the heating temperature and density of the target disc material. This means highly porous disc samples can be obtained at a much higher sintering temperature, which is usually not possible since higher hot-pressing temperature always result in fully dense samples under a certain pressure. High sintering temperatures favor coarsening of grain size due to grain growth, which would be beneficial for charge carrier transport in the TE materials; meanwhile, the as-formed porous material can scatter medium- to long-wavelength phonons, leading to a reduced lattice thermal conductivity, and potential enhancement on $ZT$. Indeed, this approach leads to a general enhancement of $ZT$s for the material systems described in this work, except of the p-type $Bi_{0.5}Sb_{1.5}Te_3$. In particular, we obtained a record high $(ZT)_{eng}$ of ~1.0, with $ZT$ ranging from ~0.9 to ~1.6 at a temperature range of 323 to 773 K, making it a promising candidate for the substitution of commercial $Bi_2Te_{2.7}Se_{0.3}$ in near room temperature applications. We studied four benchmark material systems by tuning the porosity in each of them: n-type Zintl $Mg_{3.225}Mn_{0.025}Sb_{1.5}Bi_{0.49}Te_{0.01}$, p-type $Bi_{0.5}Sb_{1.5}Te_3$, n-type half-Heuslers $Hf_{0.25}Zr_{0.75}NiSn_{0.99}Sb_{0.01}$ and p-type $Nb_{0.56}V_{0.24}Ti_{0.2}FeSb$. As the representative, the results for $Mg_{3.225}Mn_{0.025}Sb_{1.5}Bi_{0.49}Te_{0.01}$ are shown in the main text, and the others can be found in the Supporting Information (Figure S1–S3).

N-type $Mg_3Sb_2$-based materials have been recently identified as promising thermoelectric materials with $ZT$s ranging from ~0.5 to ~1.6 for temperature ranging from 300 to 773 K in $Mg_{3.0}Sb_{1.48}Bi_{0.48}Te_{0.04}$.[9] So far, intense efforts have been devoted to this material on unveiling the feature of favorable multiple band valleys at the conduction band edges,[18] tuning the extra Mg and Te (or Se) doping concentration,[19–23] and exploring the dominated carrier scattering mechanisms to boost the electrical conductivity for better $ZT$s.[24–26] Interestingly, a $ZT$ of ~0.9 at near 325 K was predicted under the modeling of the grain-boundary dominated charge transport,[27] but lack of experimental demonstration yet.

In this work, we applied the PM approach to our previously optimized composition $Mg_{3.225}Mn_{0.025}Sb_{1.5}Bi_{0.49}Te_{0.01}$ with Mn doping and extra Mg.[28] As a typical process, we first synthesized the fully dense sample both at 873 K and 1073 K using SPS under dynamic vacuum conditions, the fully dense disc pressed at 873 K were used as the precursor to make the following porous samples (Samples hot pressed at 1073 K was not used for this purpose since Mg loss at 1073 K is too much). The results of the enhanced $ZT$s are shown in Figure 1B and 1C. As shown in Figure 1B, the optimum sample with a relative density of 90% exhibits significantly high *ZT*s ranging from ~0.9 to ~1.6 at between 323 and 723 K. For comparison, the temperature dependence of *ZT*s of n-type BiTeSe materials from the literature was also shown together.[29–32] It can be seen that the PM $Mg_{3.225}Mn_{0.025}Sb_{1.5}Bi_{0.49}Te_{0.01}$ outperforms the best of $Bi_2Te_{2.7}Se_{0.3}$.

Figure 1C shows the comparison of *ZT*s in the temperature range of 300-623 K between the fully dense and 90% PM $Mg_{3.225}Mn_{0.025}Sb_{1.5}Bi_{0.49}Te_{0.01}$ samples pressed at 1073 K to other representative fully dense n-type $Mg_3Sb_2$ based materials, which were synthesized through different approaches. As shown in the figure, the *ZT* of the PM sample have been substantially boosted around the room temperature region, approaching ~0.9 based on a grain boundary dominated charge carrier transport model for this system.[27] Meanwhile, high performance at elevated temperatures was properly secured, as seen in the fully dense structure.[33,34] Overall, the global *ZT* of the PM $Mg_{3.225}Mn_{0.025}Sb_{1.5}Bi_{0.49}Te_{0.01}$ sample at 300-773 K are better than those reported thus far for n-type $Mg_3Sb_2$ based materials, and any other materials around room temperature based on the engineering figure of merit $(ZT)_{eng}$ that can reliably predict the maximum thermoelectric efficiency $\eta_{max}$ of materials at a large temperature difference. It is worth pointing out that the high performances of n-type thermoelectric materials are indeed scarcer than the p-type, and the discovery of PM n-type $Mg_3Sb_2$ based materials is expected to have a sound impact on the thermoelectric field.

To unveil the deep reasons regarding the structure and compositions underlying the high TE performance of the PM materials $Mg_{3.225}Mn_{0.025}Sb_{1.5}Bi_{0.49}Te_{0.01}$, we conducted Scanning Electron Microscope (SEM) and High Resolution Transmission Electron Microscope (HRTEM) investigations for a fully dense sample pressed at 873 K (shown in Figure 2A through 2C) and a typical PM sample with 90% relative density (shown in Figure 2D through 2I), respectively. As shown in Figure 2A, the grain size of the fully dense sample is only about 200 nm. The high-resolution HAADF image in Figure 2B suggests the existence of the spinodal decomposition of $Mg_3Sb_{1.5}Bi_{0.5}$ in the darker contrast areas, which is consistent with previous work.[28] The selected area electron diffraction (SAED) pattern shown in Figure 2C can be firmly indexed by the [001] zone axis of the α-$Mg_3Sb_2$ phase (space group No. 164, $P\bar{3}m1$). For the PM sample, substantial changes can be observed. Figure 2D shows a low-magnification SEM image of the sample surface after polishing displaying pores around a few hundred nanometers. A series of samples with decreasing relative densities shows a trend in increase of pores, as can be seen in the Supporting Information (Figure S4), indicating the effectiveness of the approach in manipulating the density and porous structures. The average grain size of the PM sample sintered at 1073 K has grown to about tens of micrometers, as shown in Figure 2E. The prominent grain growth changed the crystal defects, and boosted the electrical transport properties in the PM materials, which is similar with the previous report.[33]

Figure 2F shows the unique microstructures in the PM $Mg_{3.225}Mn_{0.025}Sb_{1.5}Bi_{0.49}Te_{0.01}$ sample featuring porridge-like nano-precipitates. High-resolution Scanning Transmission Electron Microscope (STEM) images focusing on two separate precipitates are shown in Figure 2G and 2H. The particle size of these precipitates is around 10 nm, and the crystal structure is obviously different from the host α-$Mg_3Sb_2$ phase. The Fast Fourier Transform (FFT) images (shown in the insets in Figure 2G and 2H) and the measured lattice spacing show that the precipitate in Figure 2G can be indexed to Bi with the space group of $R\bar{3}m$, and the other image in Figure 2H is Bi with a space group of $Pm\bar{3}m$. Both Bi nano-precipitates are endotaxially grown in the $Mg_3Sb_2$ based matrix, whereas the crystal lattice of first precipitate in Figure 2G is coherent to the host matrix, and for the second precipitate in Figure 2H, a clear amorphous layer of 1–2 nm in thickness can be observed between the matrix and precipitate phases, indicating the large mismatch of their lattice. The coexistence of two phases of Bi may originate from their subtle difference on the formation energy during the cooling process. In addition to the Bi nano-precipitates, the spinodal decomposition of $Mg_3Sb_{1.5}Bi_{0.5}$ exists in the darker contrast areas in Figure 2G and 2H.

It can be seen that the Bi precipitate phase exists in all PM samples. Rationally, the disassociated Bi is also accompanied by excessive Mg in the PM $Mg_{3.225}Mn_{0.025}Sb_{1.5}Bi_{0.49}Te_{0.01}$ materials, as indicated in the dark area in Figure 2I. The EDX mapping shown in the inset of Figure 2I confirms the existence of Mg pieces. We believe that the porous structures in $Mg_{3.225}Mn_{0.025}Sb_{1.5}Bi_{0.49}Te_{0.01}$ materials play a unique role in the formation of massive Bi nano-precipitates. These *in situ* formed precipitates, together with the multi-scale pores, are expected to have a profound impact on the thermoelectric transport.

On the other hand, owing to the massive existence of the Bi precipitates, concerns regarding the severe decomposition of $Mg_{3.225}Mn_{0.025}Sb_{1.5}Bi_{0.49}Te_{0.01}$ may arise, which would lead to a deterioration of its high thermoelectric performance. To further investigate the complex phases in these samples, powder neutron diffraction measurements were carried out using GPPD at China Spallation Neutron Source for both the fully dense sample annealed at 873 K, and the 90% PM $Mg_{3.225}Mn_{0.025}Sb_{1.5}Bi_{0.49}Te_{0.01}$ sample. As shown in the neutron diffraction and Rietveld refinements (shown in Figure S5), there are no detectable phases except of the pure phase patterns of the fully dense and PM samples of $Mg_{3.225}Mn_{0.025}Sb_{1.5}Bi_{0.49}Te_{0.01}$, indicating that the Bi precipitates are a minor portion in these samples. The comparison of the experimental pair distribution function G(r) (shown in Figure

S6) shows that there are no discernible differences in crystalline structure between the fully dense and PM samples.

As mentioned above, the multi-scale hierarchical structure will have a profound impact on the thermoelectric transport in PM $Mg_{3.225}Mn_{0.025}Sb_{1.5}Bi_{0.49}Te_{0.01}$ materials. Benefiting from multi-scale phonon scattering centers, particularly the massive precipitates, thermal conductivity of PM $Mg_{3.225}Mn_{0.025}Sb_{1.5}Bi_{0.49}Te_{0.01}$ materials is further reduced. The $T$-dependent $\kappa_{tot}$ for all $Mg_{3.225}Mn_{0.025}Sb_{1.5}Bi_{0.49}Te_{0.01}$ samples is shown in Figure 3A. Similar to the trend in the temperature dependent $D$ for all samples (Figure S7A), the $\kappa_{tot}$ of fully dense sample pressed at 1073K ranges from ~1.18 to ~0.85 $Wm^{-1}K^{-1}$ at 323-723 K, which are significantly higher values than those of the PM samples. As expected, the $T$-dependent $\kappa_{tot}$ of the PM samples decrease to a lower level with a decrease in the sample densities. The differences in $\kappa_{tot}$ among the PM samples within the temperature range of 323-450 K become more obvious by taking the density factor into account. As for an optimum sample of PM 90%, $\kappa_{tot}$ was effectively reduced by around 29% within the entire measured temperature range, which ranges from ~0.84 to ~0.68 $Wm^{-1}K^{-1}$ at 323-723 K. Interestingly, an identically low $\kappa_{tot}$ can be found for a nominal composition of $Mg_3Sb_{1.48}Bi_{0.48}Te_{0.04}$ in Zhang's report,[9] in which a similar processing temperature of 1123 K was used. Because no detailed microstructure or a TEM analysis was used, we presumed that the low $\kappa_{tot}$ in $Mg_3Sb_{1.48}Bi_{0.48}Te_{0.04}$ occurred owing to the reduced Mg in their sample. The actual compositions for all samples are shown in the Supporting Information (Table S1). Considering that the differences in actual Mg for all samples are minor, it is suggested the significant reduction of $\kappa_{tot}$ for the PM samples is due to unique phonon scattering mechanisms, primarily of the massive Bi precipitates, and the porosity. A phonon relaxation approximation based on the Callaway model was used to understand the details of the phonon scattering mechanisms in the PM 90% sample, as shown in the Supporting Information (Figure S7).

In addition to the highly favorable low $\kappa_L$ and $\kappa_{tot}$, the large grain size (≥20 μm) for the samples pressed at 1073 K is beneficial for the electrical transport, particularly within the low temperature range (323-473 K). The $T$ dependence of $\sigma$ for all samples is shown in Figure 3B. The value of $\sigma$ of the fully dense sample sintered at 873 K shows two distinctly different $T$-dependent regions, which is a typical characteristic for n-type $Mg_3Sb_2$ based materials synthesized at relatively lower temperatures.[24,25] It has been suggested that, within the lower temperature range, the electron transport is dominated by ionized impurity scattering, whereas at high temperatures it is dominated by acoustic phonon scattering. Another reasonable interpretation proposed by Snyder *et al.* is that the energy barriers at the crystal interfaces can dominate the electron transport, and the highest carrier mobility can be realized at the non-grain boundary limit.[27] For the PM samples and the fully dense sample sintered at 1073 K, the charge carrier scattering mechanism within the entire temperature range of 323-723 K is likely dominated by acoustic phonons with a power function of $\sigma$-$T^{-0.5}$. A distinct conversion of the dominated electron scattering mechanisms can be observed at lower temperatures between samples pressed at 873 K and 1073 K owing to significantly increased electrical mobility in the high temperature pressed samples (Figure 4A). Meanwhile, $\sigma$ of the fully dense sample (pressed at 873 K) at 323 K increased from ~2.6 × 10$^4$ S m$^{-1}$ to ~5.7 × 10$^4$ S m$^{-1}$ for the fully dense sample pressed at 1073K, and slightly reduced to ~5.2 × 10$^4$ S m$^{-1}$ for the PM Mg$_{3.225}$Mn$_{0.025}$Sb$_{1.5}$Bi$_{0.49}$Te$_{0.01}$ sample with a relative density of 90%. A room-temperature Hall measurement was carried out. As shown in Figure 4A, with the carrier concentrations dropping significantly from ~6.2 × 10$^{19}$ for the fully dense sample pressed at 873 K to ~3.7 × 10$^{19}$ cm$^{-3}$ for the fully dense sample pressed at 1073 K, their corresponding carrier mobility is prominently enhanced from ~25 to ~158 cm$^2$ V$^{-1}$ s$^{-2}$. As the density goes down, the carrier mobility decrease, whereas the carrier mobility of ~110 cm$^2$ V$^{-1}$ s$^{-2}$ for the PM 85% sample is large enough, which is responsible for the significantly enhanced electrical conductivity in the PM samples having a relative density of 95%, 90%, and 85% at 323-473 K. For the PM 80% density sample, the poor electrical conductivity can be ascribed to the low carrier concentration of ~1.5 × 10$^{19}$ cm$^{-3}$. It is noteworthy that the electrical conductivity of all PM samples at high temperature decreases compared to that of the fully dense samples, which is due to the acoustic phonon scattering of the electrons.

The $T$-dependent $S$ is shown in Figure 3C for all Mg$_{3.225}$Mn$_{0.025}$Sb$_{1.5}$Bi$_{0.49}$Te$_{0.01}$ samples. Overall, the absolute $S$ values are within the range of 200-300 μV K$^{-1}$ at 323-723 K, which are consistent with the transitional metal doped samples, but are slightly higher than those doped only by Te.[24] At 323 K, $S$ of the fully dense samples and the PM 90% density sample is around 210 μV K$^{-1}$, which is related to their relatively higher carrier concentration in comparison to other PM samples. At 373-723 K, the $S$ values decrease with the density of the samples except for the abnormally high values in the PM 80% density sample. To further understand the mechanisms of electrical transport in the PM samples, the log |$S$| - log$\sigma$ relation is shown Figure 4C. The curves predicted mainly by the two-phase grain boundary model, as well as the commonly used acoustic-phonon scattering and ionized-impurity scattering models, were compared with the experimental data points from our PM samples. Unlike the previous reports for fully dense Mg$_3$Sb$_2$ based materials, whose data fit fairly well with either of three models,[27,34] the experiment data in this work are located in the upper-right region of the panel in Figure 4C, which exceeds the predictions of the models based on the grain boundary scattering assumptions, indicating superior electrical transport properties in these $Mg_{3.225}Mn_{0.025}Sb_{1.5}Bi_{0.49}Te_{0.01}$ samples. Similarly, we did observe high electrical conductivity and Seebeck coefficient at near 323 K for the nominal $Mg_{3.175}Mn_{0.025}Sb_{1.5}Bi_{0.49}Te_{0.01}$ sample pressed at 873 K in our previous report,[28] which also exceeds the predictions of the above electron transport models. In the current work, we particularly increased the Mg content by 1.6% in the synthesis of precursor disc samples at 873 K to avoid further loss of Mg during the second SPS process. As can be seen that the obtained fully dense sample exhibited as high carrier concentration as $6.2 \times 10^{19}$ cm$^{-3}$, in comparison to $3.0 \times 10^{19}$ cm$^{-3}$ for the $Mg_{3.175}Mn_{0.025}Sb_{1.5}Bi_{0.49}Te_{0.01}$ sample, meanwhile the carrier mobility reduced by 50% through adding excess Mg, leading to similar electrical conductivity. Accordingly, the seebeck coefficient at near 323 K reduced from 220 to 210 μV K$^{-1}$ with the increase of carrier concentration in $Mg_{3.225}Mn_{0.025}Sb_{1.5}Bi_{0.49}Te_{0.01}$, which is higher than the seebeck values for the similar compositions without Mn doping[23,33]. Indeed, as has been reported.[28] Mn doping in $Mg_3Sb_2$ based materials exhibits higher effective mass and power factors. The superior electrical transport properties in the precursor sample pressed at 873 K were succeeded by the full density and PM samples pressed at 1073 K, which outperforms the predicted values in the curve based on the two-phase grain boundary model at the no-grain boundary limit.[27]

Combining the drastically reduced thermal conductivity within a broad temperature range and the promoted power factors (Figure 4B) at near room temperatures, the *ZT*s of the $Mg_{3.225}Mn_{0.025}Sb_{1.5}Bi_{0.49}Te_{0.01}$ materials at near room temperature have been greatly improved, as shown in Figure 3D. In particular, the *ZT* of the 90% PM sample reaches ~0.88 at 323 K, which is an increase of 140% compared to that of the fully dense sample pressed at 873 K, and 25% to that of the fully dense sample pressed at 1073K. The highest peak *ZT* of which is ~1.75, is obtained for the 95% PM sample at 723 K. Moreover, the *ZT*s of other PM samples are also significantly enhanced at near room temperature compared to fully dense $Mg_{3.225}Mn_{0.025}Sb_{1.5}Bi_{0.49}Te_{0.01}$. In addition, the repeatability of the thermoelectric properties in the $Mg_3Sb_2$ based PM materials and the reproducibility of the PM 90% samples are conformed, as shown in the Supporting Information (Figure S8).

To prove the significant potential of a porous-mediated structure in achieving high-performance thermoelectrics, we further extended the rationally designed porous structure to three other benchmark TE materials, namely, $Bi_{0.5}Sb_{1.5}Te_3$, $Hf_{0.25}Zr_{0.75}NiSn_{0.99}Sb_{0.01}$, and $FeNb_{0.56}V_{0.24}Ti_{0.2}Sb$. Notable enhancement of thermoelectric performance was realized in half-Heuslers. However, a decayed performance was seen from $Bi_{0.5}Sb_{1.5}Te_3$ by introducing porosity, which is ascribed to the less gains on lattice thermal conduction reduction comparing other porous materials described in this work. The detailed thermoelectric properties for these samples can be found in the Supporting Information. As the key parameter, the average *ZT*s of these materials is shown in Figure 5. A prominent improvement with respect to the fully dense material has been realized for three systems, and thus we expect that the PM structure can also be extended to more state-of-art the material systems for an enhanced thermoelectric performance.

In summary, we developed a porous-mediated (PM) approach for achieving controllable porosity and nano precipitates for high thermoelectric performance. We applied this approach to the $Mg_3Sb_2$ based n-type thermoelectric materials $Mg_{3.225}Mn_{0.025}Sb_{1.5}Bi_{0.49}Te_{0.01}$, demonstrating *ZT*s of ~0.9 and ~1.6 at 323 and 723 K, respectively, making it competitive with BiTeSe for near room temperature applications. The high performance was ascribed to the enhanced electrical mobility and multi-scale phonon scattering centers, particularly for the well-dispersed bismuth nano-precipitates induced through the porous structure. We further extended the PM approach to other thermoelectric materials including half-Heuslers $FeNb_{0.56}V_{0.24}Ti_{0.2}Sb$ and $Hf_{0.25}Zr_{0.75}NiSn_{0.99}Sb_{0.01}$, with enhanced properties, paving the way for high-performance PM thermoelectrics.

**Experimental Section**

*Materials*: Magnesium turnings (Mg, 99.98%; Alfa Aesar), bismuth pieces (Bi, 99.99%; Alfa Aesar), antimony shots (Sb, 99.8%; Alfa Aesar), tellurium pieces (Te, 99.999%; Alfa Aesar), manganese powder (Mn, 99.99%; Alfa Aesar), iron shots (Fe, 99.98%; Alfa Aesar), titanium shots (Ti, 99.98%; Alfa Aesar), niobium shots (Nb, 99.98%; Alfa Aesar), vanadium shots (V, 99.8%; Alfa Aesar), hafnium shots (Hf, 99.8%; Alfa Aesar), zirconium shots (Zr, 99.8%; Alfa Aesar), nickel shots (Ni, 99.8%; Alfa Aesar), and tin shots (Sn, 99.99%; Alfa Aesar) were used.

*Synthesis*: The Schematic is as shown in Figure 1A in the article. The process uses a specially designed graphite die with a height of 40 mm, and the total length of the top and bottom graphite rods plus the thickness (2–3 mm) of a fully dense disc material was set to be equal to the height of the graphite die. Thus, a constant volume space between two rods can be left in the die to modulate the ratio of porosity (or relative density) of the final samples by simply tuning the quantity of the pre-ball milled fine powder.

For the synthesis of the PM $Mg_{3.225}Mn_{0.025}Sb_{1.5}Bi_{0.49}Te_{0.01}$ materials, the raw elements were weighed according to the composition of $Mg_{3.225}Mn_{0.025}Sb_{1.5}Bi_{0.49}Te_{0.01}$. The elements were loaded into a stainless-steel ball-milling jar in a glove box under an argon atmosphere with an oxygen level of below 0.1 ppm and ball-milled for 10 h using a SPEX 8000M Mixer/Mill. Then, for the initial SPS, a regular graphite die loaded with 5.00 g of reactant powder was immediately sintered at 873 K under a pressure of ~50 MPa for 2 min. Next, the fully dense ingot with a thickness of ~5 mm obtained was smashed and loaded into another clean ball-milling jar, followed by a 6 h ball milling process. For the synthesis of the PM samples, the powder obtained was then loaded into our specially designed graphite die with a constant inner volume size for use in another SPS process at 1073 K under dynamic vacuum conditions. During the synthesis of the PM samples, a series of relative densities can be reached by fine tuning the quantity of the powder; however, we should note that, as the relative density reaches above 95%, the samples can be squeezed out through melting, and thus the maximum density was 95% for the synthesis.

For $FeNb_{0.56}V_{0.24}Ti_{0.2}Sb$, the constituent elements were ball-milled for 10 h, and then SPS-sintered at 1073 K under a pressure of 50 MPa for 10 min. The ingot obtained was characterized by XRD to be in a pure half-Heusler phase. For the PM samples, the ingot obtained was ball-milled for 6 h, and SPS-sintered for a second time at 1173 K for 10 min using a specially designed graphite die.

For the $Hf_{0.25}Zr_{0.75}NiSn_{0.99}Sb_{0.01}$ ingot, instead of an initial SPS, the fully dense ingot was prepared through arc melting, followed by a 10 h ball-milling process, and then sintered at 1173 K for 20 min using the specially designed graphite die.

For PM $Bi_{0.5}Sb_{1.5}Te_3$, the constituent elements were ball-milled for 10 h, and then sintered at 723 K under a pressure of 50 MPa for 10 min by directly applying the specially designed graphite die without a second SPS.

*Electrical transport properties*: Bar samples were cut from the pressed disks with dimensions of ~10 mm × 2 mm × 2 mm and used for the simultaneous measurement of the electrical resistivity ($\rho$) and Seebeck coefficient ($S$) on a commercial system (Linseis LSR-3, Germany) under a helium atmosphere.

*Hall measurements*: The samples obtained were cut into slices with dimensions of ~5 mm × 5 mm × 0.2 mm and then welted for Hall measurements. The room-temperature Hall coefficient ($R_H$) was measured under a reversible magnetic field using a physical properties measurement system (PPMS). The Hall carrier concentration ($n_H$) was obtained using $n_H = 1/eR_H$, and the Hall carrier mobility ($\mu_H$) was calculated by applying $\sigma = e\mu_H n_H$, where $e$ is the electronic charge and $\sigma$ is the electrical conductivity.

*Thermal conductivity*: The thermal diffusivity coefficient ($D$) was measured on a laser flash system (Linseis LFA 1000 Laser Flash, Germany). The specific heat capacity ($C_p$) was measured on a differential scanning calorimetry thermal analyzer (Netzsch DSC 404 C, Germany). The total thermal conductivity $\kappa_{tot}$ is calculated from $\kappa_{tot} = DC_P d$, where $D$ is the thermal diffusivity coefficient, $C_P$ is the specific heat capacity, and $d$ is the density.

*Aberration-corrected transmission electron microscopy (Cs-corrected TEM)*: The samples investigated through transmission electron microscopy (TEM) were prepared using traditional mechanical polishing, dimpling, and ion milling with liquid nitrogen. High-angle annular dark-field (HAADF) imaging was carried out using a JEOL ARM 200 F microscope, equipped with probe and image correctors.

*Chemical composition analysis*: The chemical compositions were analyzed using ICP-MS (X Series 2, Thermo Fisher Scientific).

*Neutron diffraction measurements*: Powder neutron diffraction patterns are obtained from GPPD (90° bank) at China Spallation Neutron Source in China at room temperature and the Rietveld refinement is conducted using the Z-Rietveld program. Neutron total scattering measurements were carried out using the high-intensity total diffractometer beamline BL21 NOVA of J-PARC in Japan at room temperature under a user program (Proposal No. 2018A0286). The background, attenuation factor, number of incident neutrons, solid angle of the detectors, multiple scattering, and incoherent scattering cross-sections were corrected, and the intensity was normalized to determine the static structure factor $S(Q)$. The PDF data were calculated through a Fourier transform of $S(Q)$ with a cutoff ($Q_{max}$) of 30 Å$^{-1}$. The same parameters were applied to the data for both samples. The real-space refinement of experimental $G(r)$ is performed using the PDFgui program.

**Supporting Information**
Supporting Information is available from the Wiley Online Library or from the author.

**Acknowledgements**
The authors acknowledge the funding support from the National Natural Science Foundation of China (NSFC) under the grant numbers of 51572287, 11675255 and U1601213. The project is also supported by the Key Program of Frontier Sciences of the Chinese Academy of Sciences (Grant No. QYZDB-SSW-SLH013). Xiaoxi Chen and Siyi Chang contributed equally to this work.

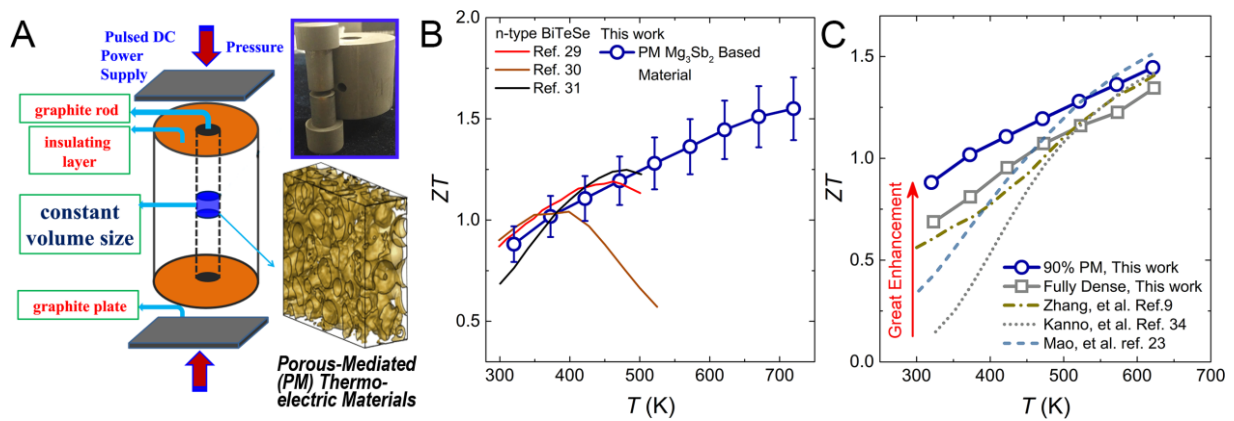

**Figure 1.** (A) Schematic illustration of the synthesis of high-performance porous-mediated (PM) thermoelectric materials. (B) Temperature-dependent figure-of-merit *ZT* of the PM $Mg_{3.225}Mn_{0.025}Sb_{1.5}Bi_{0.49}Te_{0.01}$ material with relative density of 90%, in comparison to the state-of-the-art n-type BiTeSe materials in the literature. (C) Comparison of the temperature-dependent *ZT* values between PM and full density $Mg_{3.225}Mn_{0.025}Sb_{1.5}Bi_{0.49}Te_{0.01}$ materials pressed at 1073K in this work and the reported results from the literature.

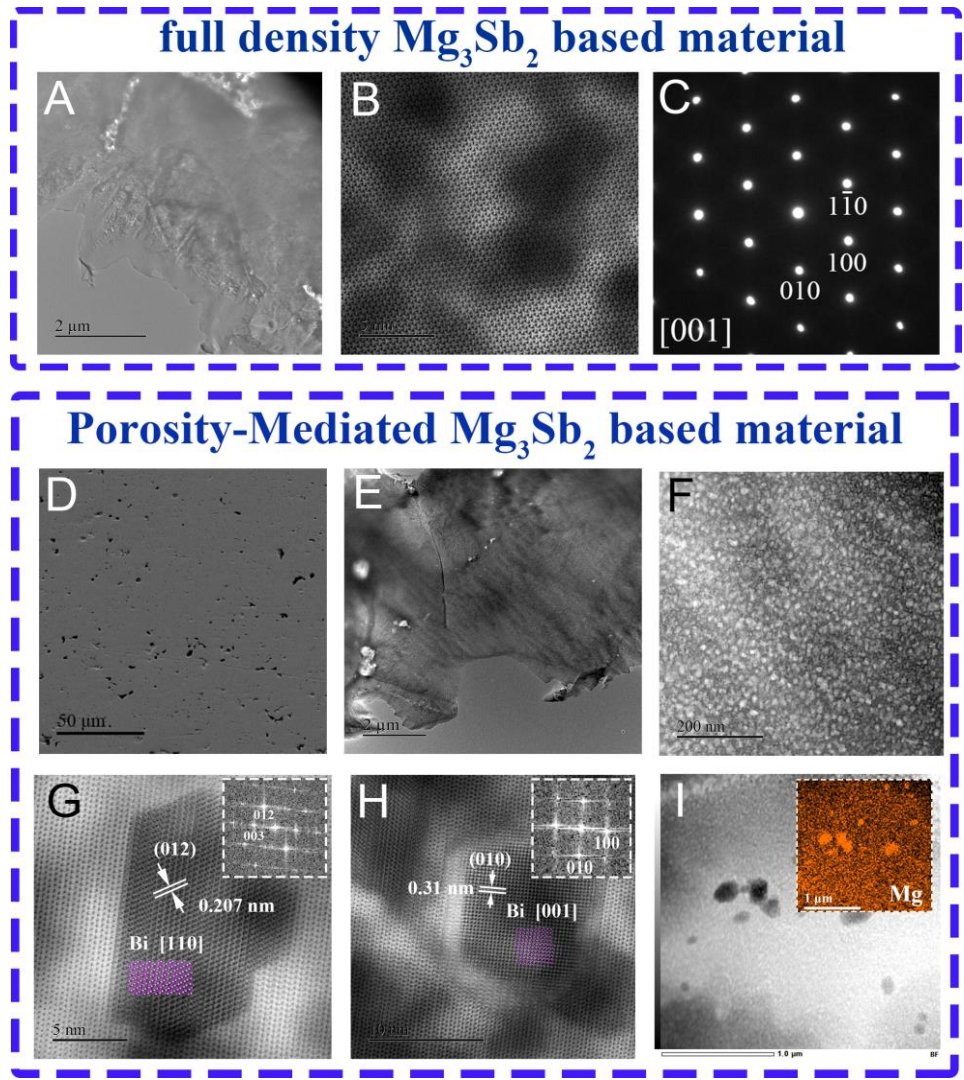

**Figure 2.** Hierarchical microstructure of the fully dense (SPS at 873K, A to C) and 90% relatively dense PM samples (D to I) of $Mg_{3.225}Mn_{0.025}Sb_{1.5}Bi_{0.49}Te_{0.01}$. For the fully dense sample: (A) Low magnification SEM image. (B) HRTEM image of the lattice matrix. (C) Selected area electron diffraction (SAED) pattern along the zone axis [001]. For the 90% PM sample: (D) Low-magnification SEM image after the surface polishing process. (E) Low-magnification TEM image. (F) Porridge-like distribution of Bi precipitates. (G), (H) Two different indexed structures of Bi precipitates under HRTEM, the insets of which are the corresponding FFT images. (I) Selected TEM area with excess Mg precipitates, the inset of which shows EDX mapping.

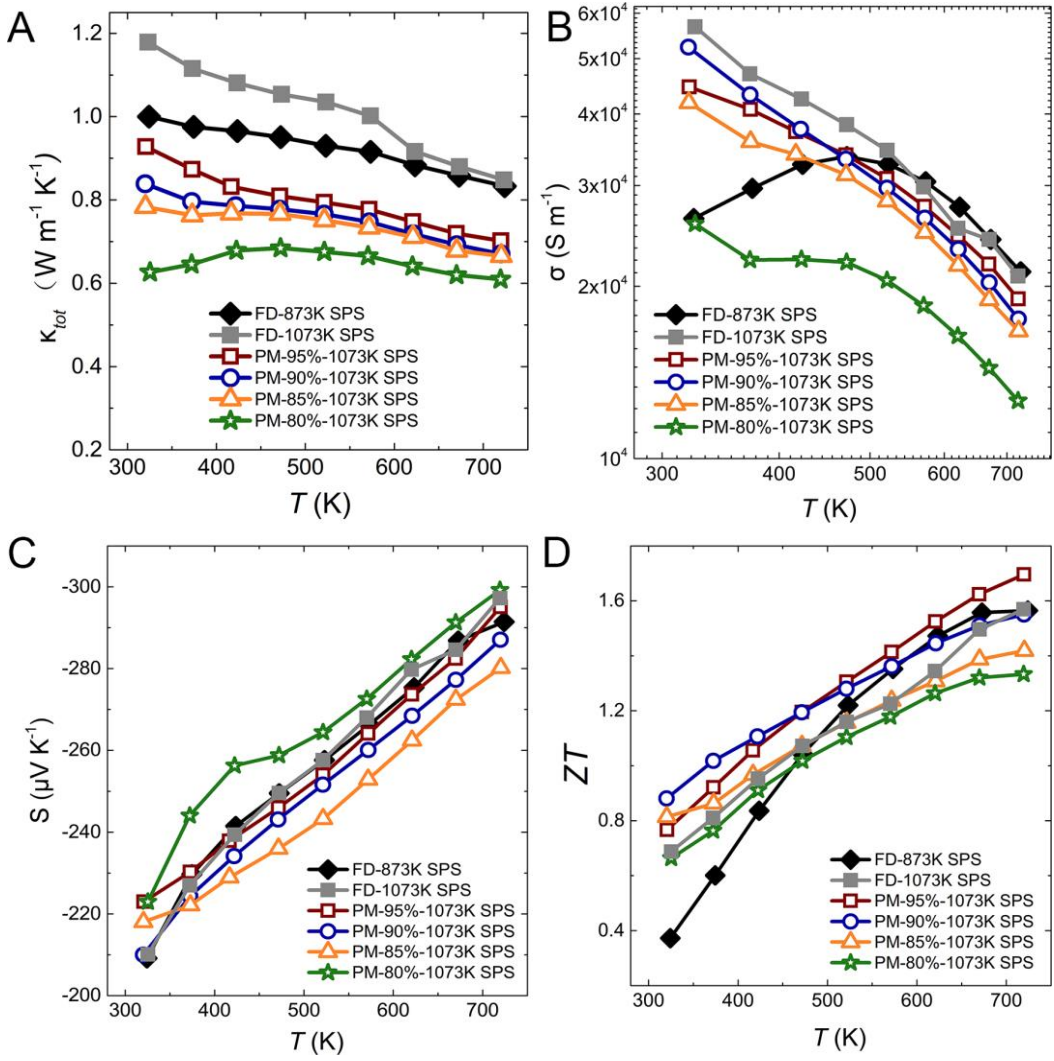

**Figure 3.** Thermoelectric properties of a series of PM $Mg_{3.225}Mn_{0.025}Sb_{1.5}Bi_{0.49}Te_{0.01}$ materials. (A) total thermal conductivity. (B) $T$ dependence of $\sigma$. (C) $T$ dependence of $S$. (D) $T$ dependence of $ZT$ values for the fully dense and PM samples.

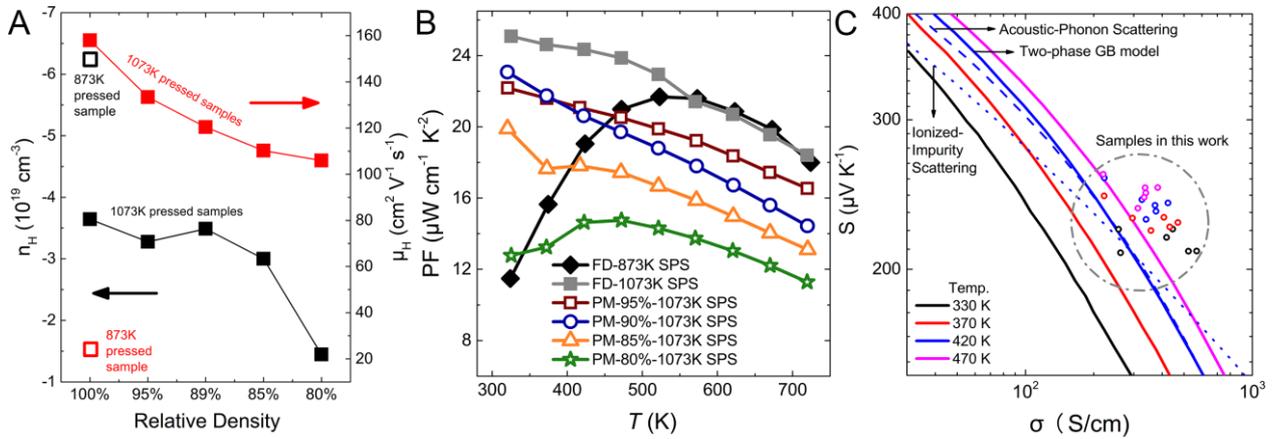

**Figure 4.** Thermoelectric properties of a series of PM $Mg_{3.225}Mn_{0.025}Sb_{1.5}Bi_{0.49}Te_{0.01}$ materials. (A) carrier concentration ($n_H$) and mobility ($\mu_H$) for a series of fully dense and PM samples.

(B) *T* dependence of *PF*. (C) The log |*S*| - log*σ* relation. The data points are from the PM samples of a series of relative densities, and the solid lines represent the two-phase grain boundary models established by Kuo *et al.*, and calculated curves at 420 K based on two other models are also presented for comparison.[49]

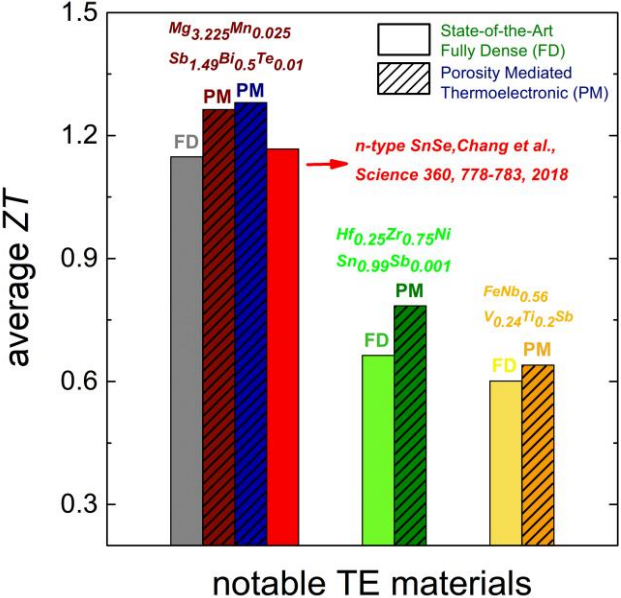

**Figure 5.** Comparison of average *ZT*s between fully dense samples and corresponding PM samples for broad extensive thermoelectric materials. The red column represents the data calculated from n-type SnSe.[7]

**A porous-mediated (PM) N-type Mg3Sb2 based material with ZT values comparable to that of the best N-type Bi$_2$Te$_{2.3}$Se$_{0.7}$ was developed**, which opens up a new class of materials for cooling and power generation at low temperature. The porosity was further extended to thermoelectric materials FeNb$_{0.56}$V$_{0.24}$Ti$_{0.2}$Sb and Hf$_{0.25}$Zr$_{0.75}$NiSn$_{0.99}$Sb$_{0.01}$, and enhanced properties are also realized, paving the way for high-performance PM thermoelectrics.

**Thermoelectric**

*Xiaoxi Chen, Siyi Chang, Jin Chen, Pengfei Nan, Xiyang Li, Xu Chen, Xiaoshan Zhu, Binghui Ge\*, Wei Cai, Jiehe Sui\*, Shuqi Zheng, Fangwei Wang, Xiaolong Chen, Huaizhou Zhao\*.*

**Porosity-mediated High-performance n-type Mg$_3$Sb$_2$ Based Thermoelectric Materials**

ToC figure

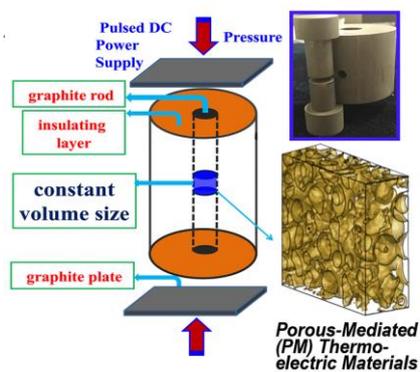

Supporting Information

**Porosity-mediated High-performance n-type Mg$_3$Sb$_2$ Based Thermoelectric Materials**

*Xiaoxi Chen, Siyi Chang, Jin Chen, Pengfei Nan, Xiyang Li, Xu Chen, Xiaoshan Zhu, Binghui Ge\*, Wei Cai, Jiehe Sui\*, Shuqi Zheng, Fangwei Wang, Xiaolong Chen, Huaizhou Zhao\*.*

**Supplementary Text**

Calculation of the maximum thermoelectric efficiency $\eta_{max}$

The maximum thermoelectric efficiency $\eta_{max}$ of materials at a large temperature difference can be calculated by using the following formula [1]:

$$\eta_{max} = \eta_c \frac{\sqrt{1+(ZT)_{eng}(\hat{\alpha}/\eta_c - 1/2)}-1}{\hat{\alpha}\left(\sqrt{1+(ZT)_{eng}(\hat{\alpha}/\eta_c - 1/2)}+1\right)-\eta_c}, \qquad [1]$$

where $\eta_c$ is the Carnot efficiency, $(ZT)_{eng}$ is the engineering dimensionless figure of merit, and $\hat{\alpha}$ is a dimensionless intensity factor of the Thomson effect, defined as $\hat{\alpha}=S(T_h)\,\Delta T/\int_{T_c}^{T_h} S(T)dT$, where $S(T_h)$ is the Seebeck coefficient at the hot-side temperature, $T_h$. The $(ZT)_{eng}$ (Figure S9) of the PM Mg$_{3.225}$Mn$_{0.025}$Sb$_{1.5}$Bi$_{0.49}$Te$_{0.01}$ sample with a relative density of 90% can be calculated based on the equations in (49).

Lattice thermal conductivity calculation and the phonon scattering mechanisms

The equations for different relaxation times become the following:

$$\tau_U^{-1} = \frac{\hbar \gamma^2 \omega^2 T}{M v_m^2 \theta_D} \exp\left(\frac{-\theta_D}{3T}\right) \qquad [1]$$

$$\tau_{EP}^{-1} = \frac{E_{def}^2 m^{*2} \omega}{2\pi \hbar^3 d v_l} \qquad [2]$$

$$\tau_{PD}^{-1} = \frac{V\Gamma \omega^4}{4\pi v_m^3} \qquad [3]$$

$$\tau_B^{-1} = \frac{v_m}{l} \qquad [4]$$

where $\gamma$ is the Grüneisen parameter, $\theta_D$ is the Debye temperature, M is the average atomic mass, $E_{def}$ is the deformation potential, m$^*$ is the density of the state effective mass, $d$ is the density, V is the volume per atom, $\Gamma$ is the disorder scattering parameter, l is the average grain size, $v_t$ is the longitudinal phonon velocity, $v_t$ is the transverse phonon velocity, and $v_m$ ($v_m = \left[\frac{1}{3}\left(\frac{2}{v_t^3} + \frac{1}{v_l^3}\right)\right]^{-1/3}$) is the average phonon velocity.

All four scattering mechanisms above were taken into account for phonon scattering modeling of the fully dense samples. For the PM Mg$_{3.225}$Mn$_{0.025}$Sb$_{1.5}$Bi$_{0.49}$Te$_{0.01}$ materials having a relative density of 95% and 90%, other scattering mechanisms will be taken into account, for example, the porosity and the precipitate nanoparticles. The relaxation time of the second-phase nanoparticles can be calculated as follows:

$$\tau_p^{-1} = v(\sigma_s^{-1} + \sigma_l^{-1})^{-1} d_p \qquad [5]$$

where d$_p$ is the density of the nanoparticles, and $\sigma_l$ and $\sigma_s$ are the cross-section limits given by

$$\sigma_s = 2\pi R^2 \qquad [6]$$

$$\sigma_l = \pi R^2 \frac{4}{9}\left(\frac{\Delta d}{d}\right)^2 \left(\frac{\omega R}{v_m}\right)^4 \qquad [7]$$

Here, $\Delta d$ is the difference in density between the nanoparticles and the matrix materials, d is the density of the matrix, and R is average radius of the sphere in the second phase.

The effects of porosity on the lattice thermal conductivity is given by

$$\frac{\kappa_p}{\kappa_b} = \frac{1-\varphi}{1+\varphi/2} \qquad [8]$$

where $\kappa_p$ and $\kappa_b$ are the lattice thermal conductivity of the porosity and bulk samples, respectively, and $\varphi$ is the porosity.

The parameters used for the modeling of the lattice thermal conductivity are listed in Table S2.

Lorenz number calculations

The Lorenz number as a function of the reduced Fermi energy $\xi = E_f/k_B T$ and scattering parameter $r$ is as follows (47):

$$L = \left(\frac{k_B}{e}\right)^2 \left(\frac{(r+7/2)F_{r+5/2}(\xi)}{(r+3/2)F_{r+1/2}(\xi)} - \left[\frac{(r+5/2)F_{r+3/2}(\xi)}{(r+3/2)F_{r+1/2}(\xi)}\right]^2\right) \qquad [9]$$

where $F_n$ is the Fermi integral given by

$$F_n(\xi) = \int_0^\infty \frac{\chi^n}{1+e^{\chi-\xi}} d\chi \qquad [10]$$

In this calculation, acoustic phonon scattering is assumed as the main carrier scattering mechanism using $r = -0.5$.

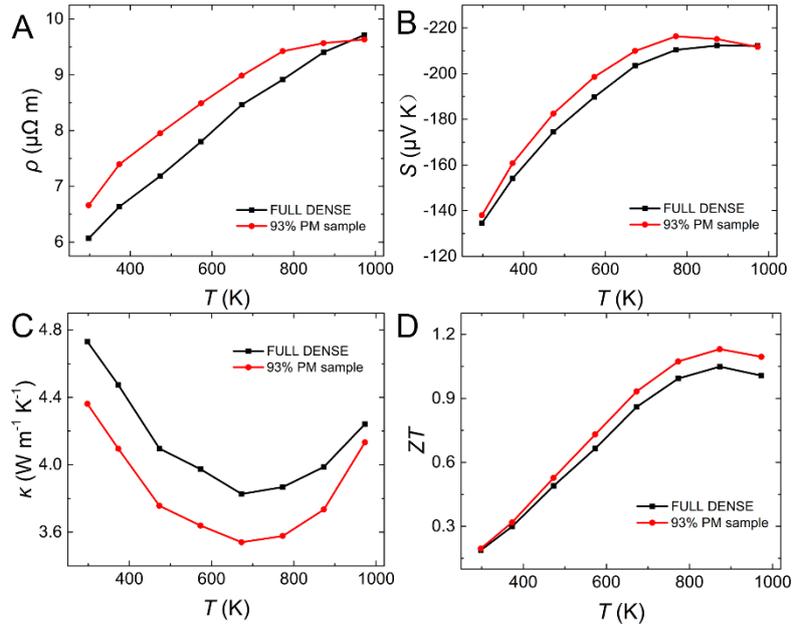

**Figure S1.**
Thermoelectric properties of the optimized PM 93% $Hf_{0.25}Zr_{0.75}NiSn_{0.99}Sb_{0.01}$ and its corresponding fully dense materials. (A) $T$ dependence of $\rho$. (B) $T$ dependence of $S$. (C) $T$ dependence of total thermal conductivity $\kappa$. (D) $T$ dependence of $ZT$.

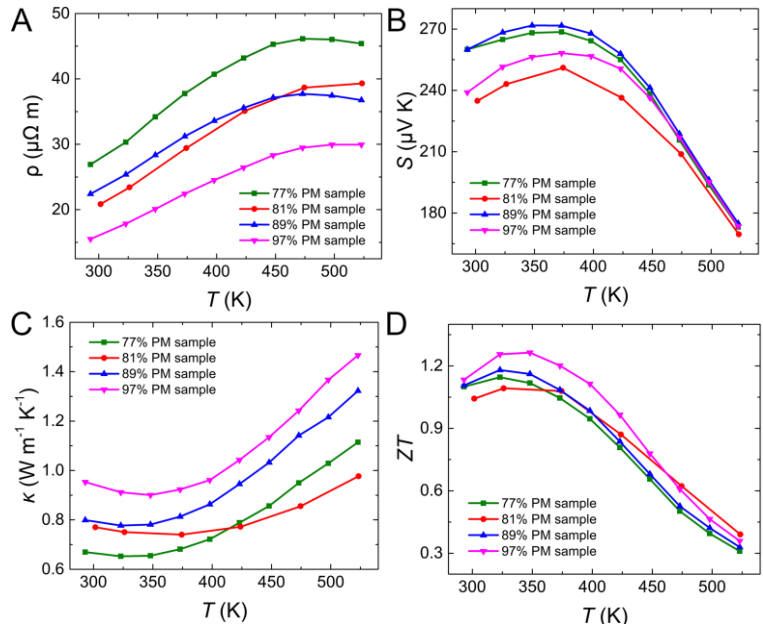

**Figure S2.**
Thermoelectric properties of the PM $Bi_{0.5}Sb_{1.5}Te_3$ and the corresponding fully dense materials. (A) $T$ dependence of $\rho$. (B) $T$ dependence of $S$. (C) $T$ dependence of total thermal conductivity $\kappa$. (D) $T$ dependence of $ZT$.

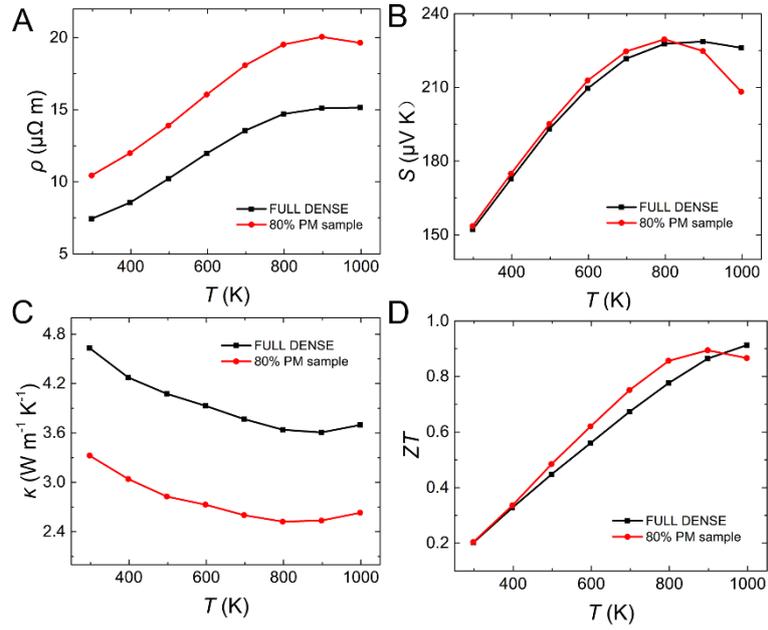

**Figure S3.**
Thermoelectric properties of the optimized PM 80% FeNb$_{0.56}$V$_{0.24}$Ti$_{0.2}$Sb and the corresponding fully dense materials. (A) $T$ dependence of $\rho$. (B) $T$ dependence of $S$. (C) $T$ dependence of total thermal conductivity $\kappa$. (D) $T$ dependence of $ZT$.

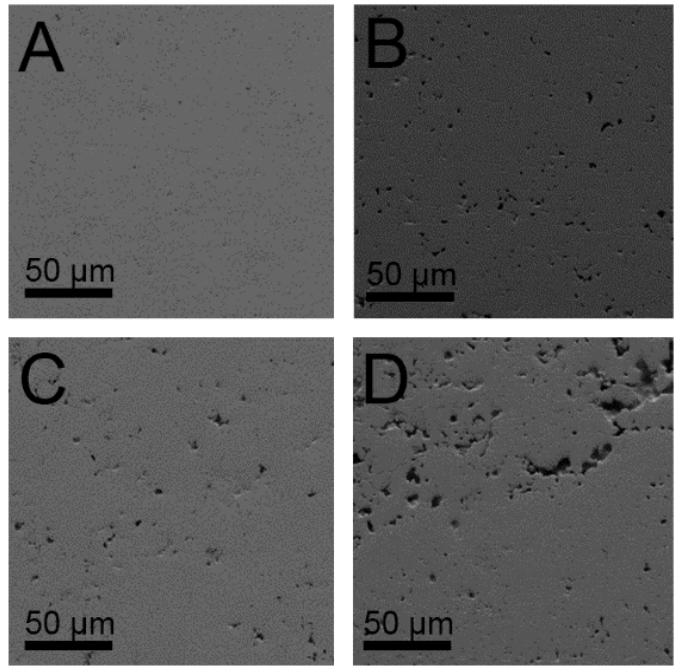

**Figure S4.**
Comparison of SEM images among $Mg_{3.225}Mn_{0.025}Sb_{1.5}Bi_{0.49}Te_{0.01}$ PM samples with a series of relative density. (A) through (D) correspond to the samples with a relative density of 95%, 90%, 85%, and 80%, respectively.

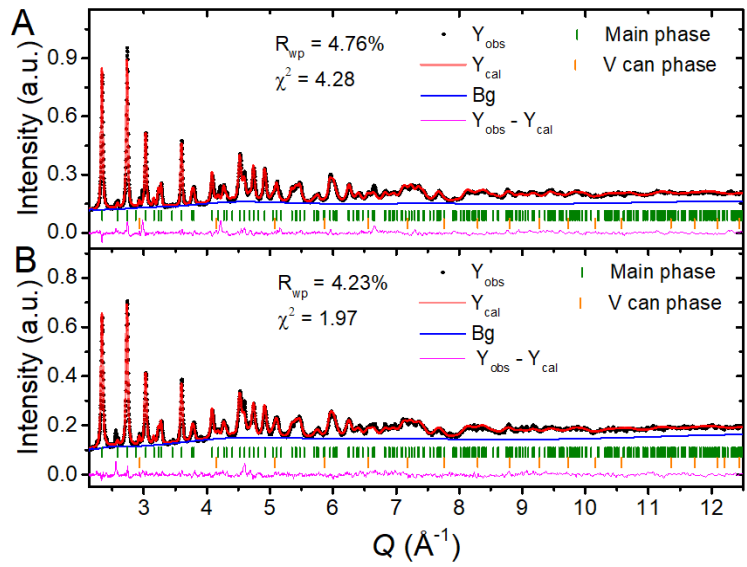

**Figure S5.**
Rietveld refinement results. (A), (B), the refinements of neutron powder diffraction patterns obtained from GPPD at China Spallation Neutron Source (90° bank) at room temperature with $P\bar{3}m1$ (space group No. 164) structure for fully dense and PM samples, respectively. A vanadium sample can phase can be verified.

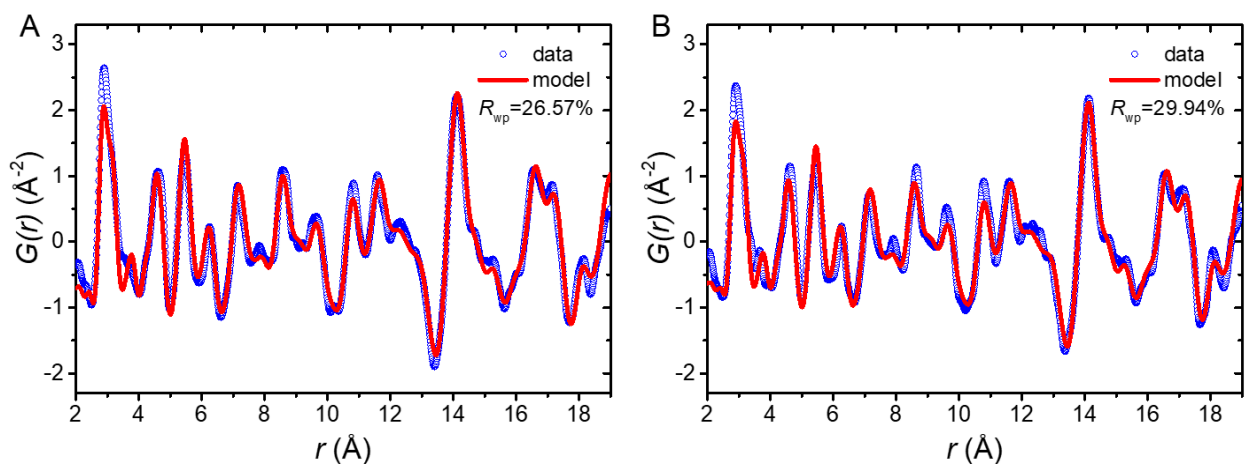

**Figure S6.**
Experimental pair distribution function $G(r)$, obtained through neutron total scattering measurement using a high intensity total diffractometer BL21 NOVA at J-PARC at room temperature for a (A) fully dense sample pressed at 873K and (B) PM sample, respectively. The real-space refinement of experimental $G(r)$ is based on the $P\bar{3}m1$ (space group No.164) structure model.

Figure S6 shows a comparison of the experimental pair distribution function, G(r), at room temperature for both the fully dense and PM samples. The results show that there are no discernible differences in crystalline structure between the fully dense and PM samples, whereas a significant difference was found for the peak located at $r$ of ~3.8 Å, which can be identified as Mg-Mg pairs in the structure. It can be seen that, for the PM sample, the experimental peak fits well with the model, whereas for the fully dense sample, the experimental peak seems weaker than that of the model that indicate more Mg vacancies. These findings are consistent with the electrical properties in both samples for which Mg vacancies play a role.

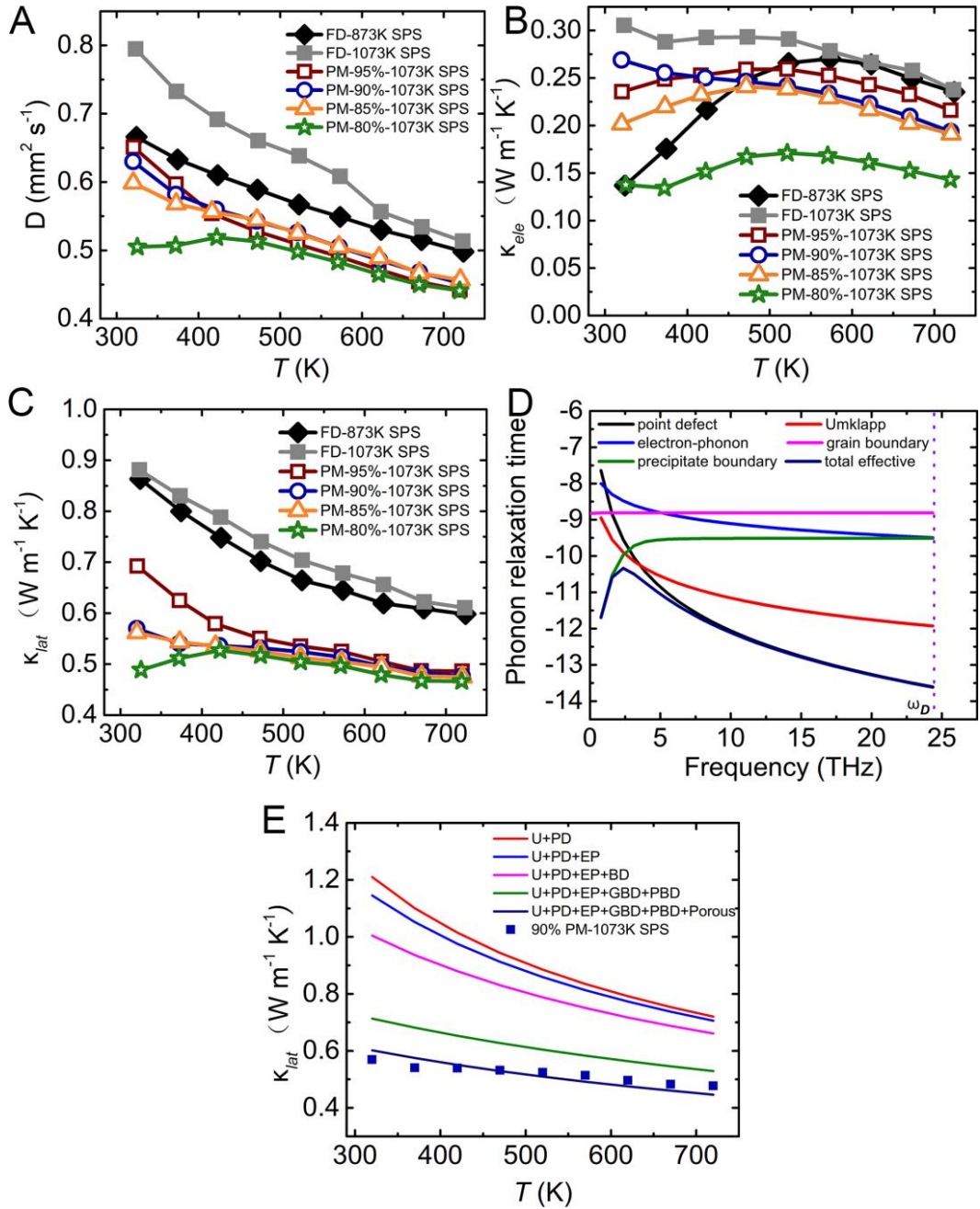

**Figure S7.**
Thermal properties as a function of temperature, and the calculated models for PM $Mg_{3.225}Mn_{0.025}Sb_{1.5}Bi_{0.49}Te_{0.01}$ materials. (A) Thermal diffusion coefficient. (B), (C) Electrical $\kappa_e$ and lattice thermal conductivity $\kappa_L$, respectively. (D) Calculated phonon relaxation time as a function of frequency for different phonon scattering mechanisms in 90% relative dense PM $Mg_{3.225}Mn_{0.025}Sb_{1.5}Bi_{0.49}Te_{0.01}$ material. (E) Fitted lattice thermal conductivity for different scattering process in PM $Mg_{3.225}Mn_{0.025}Sb_{1.5}Bi_{0.49}Te_{0.01}$ material, the experimental values of which fit well with the models.

The temperature dependence of the thermal diffusivity for a series of PM samples with the nominal composition of $Mg_{3.225}Mn_{0.025}Sb_{1.5}Bi_{0.49}Te_{0.01}$ is shown in Figs. S7A, respectively. In terms of the fully dense samples, the sample pressed at 1073K exhibit significantly higher thermal diffusion coefficient, electrical thermal conductivity than that pressed at 873K, owing to its larger grain size and high electrical conductivity in the whole temperature range.

Naturally, the relatively low density (*d*) in the PM samples would cause a reduction of $\kappa_{tot}$ according to the formulation above, which can be understood owing to the porosity factor. Beyond that, as can be seen in Figure S7A, all PM samples exhibit an evident reduction in their thermal diffusivity *D*. At 320 K, *D* drops gradually from ~0.79 mm$^2$ S$^{-1}$ for the fully dense sample (SPS at 1073 K) to 0.65 mm$^2$ S$^{-1}$ for the PM 95% sample, then to ~0.63 mm$^2$ S$^{-1}$ for the PM 90% sample, and finally ~0.50 mm$^2$ S$^{-1}$ for the PM 80% sample. At higher temperatures of up to 723 K, *D* drops from ~0.51 mm$^2$ S$^{-1}$ for the fully dense sample to ~0.45 mm$^2$ S$^{-1}$ for all PM samples. Exceptionally, for the 80% PM sample, the diffusivity near room temperature becomes abnormally low, suggesting the existence of other dominated phonon scattering mechanisms.

As shown in Figure S7B, the *T*-dependent electrical thermal conductivity $\kappa_e$ for all samples was determined using the Wiedemann-Franz law $\kappa_e = L\sigma T$, where *L*, $\sigma$, and *T* are the Lorenz number (Fermi energy dependent *L*), *T*-dependent electrical conductivity, and absolute temperature, respectively. As can be seen, $\kappa_e$ of the fully dense sample pressed at 873K shows a different feature of *T* dependence on all other samples pressed at 1073K, which is closely linked to their electric transport mechanisms.

As shown in Figure S7C. Clearly, a $\kappa_L$ gap exists between the fully dense sample and the group of PM samples. For the fully dense sample pressed at 873K, $\kappa_L$ ranges from ~0.85 to ~0.60 Wm$^{-1}$ K$^{-1}$ at 323-723 K, which matches well with most of the reported values. (*46–48*) For another fully dense sample pressed at 1073K, $\kappa_L$ increases due to the grain growth. For the PM samples, $\kappa_L$ first decreases with the density at low temperatures, and then coincides at around 500 K, indicating the various phonon scattering mechanisms among them. A typical Umklapp scattering with $\kappa_L$ of ~$T^{-1}$ can be observed for both fully dense and 95% PM samples, whereas for the PM samples other mechanisms start to dominate, which will be analyzed in the following theoretical fitting.

A phonon relaxation approximation based on the Callaway model was used to understand the details of the phonon scattering mechanisms, as shown in Figs. S7D and S7E. The details of the modeling can be found in the Supplementary Text. Herein, considering the specially designed porous structure and the novel Bi nano-precipitates, additional factors have to be taken into account for the PM samples on top of the scatterings from the Umklapp process, point defect, and so on. Based on the calculations, the frequency-dependent phonon relaxation time for the PM samples is shown in Figure S7D. The results show that the boundary phonon scattering between the Bi precipitate and host matrix is the dominant mechanism at below 2.5 THz, and Umklapp scattering is the dominant mechanism for medium- to high-frequency phonons of between 2.5 and 25 THz in PM materials. This explains the different behaviors of the *T*-dependent $\kappa_L$ in Figure S7C. At low temperature, the density of the boundaries between the Bi precipitates and the host matrix govern the phonon scattering. Therefore, the sample with a higher density of boundaries or higher density of precipitates should exhibit a lower lattice thermal conductivity, which can be easily seen in Figure S7C. For instance, the 80% PM sample exhibits the lowest $\kappa_L$ at a low *T*, which is consistent with its highest density of the precipitates. At a higher temperature, Umklapp scattering starts to dominate, and $\kappa_L$ for the PM samples behaves in quite the same manner. A comparison between the experimentally observed $\kappa_L$ and the fitted $\kappa_L$ based on the Callaway model for a PM sample of 90% relative density is shown in Figure S7E. The total fitted $\kappa_L$ when considering all mechanisms matches well with the experiment data. Among the scatterings, the boundaries between Bi precipitates and the host matrix, along with the Umklapp process, contribute the most in blocking the transmission of heat carrying phonons, leading to a substantial decrease of $\kappa_L$. As a whole, by intentionally fabricating the porous structure accompanied by massive Bi precipitates in the $Mg_{3.225}Mn_{0.025}Sb_{1.5}Bi_{0.49}Te_{0.01}$ samples, $\kappa_L$ has been unprecedentedly reduced to a low level.

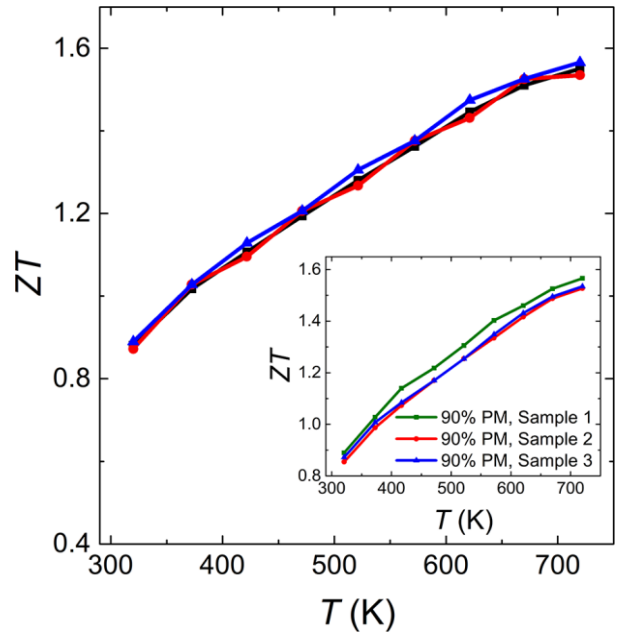

**Figure S8.**
*T*-dependent *ZT* curves based on three measurements for the same $Mg_{3.225}Mn_{0.025}Sb_{1.5}Bi_{0.49}Te_{0.01}$ sample. The inset shows the *ZT* curves for three different 90% PM samples indicating of the repeatability of the synthesis process.

Repeated tests were carried out for the PM 90% sample, and the measured *S*, $\sigma$, and $\kappa_{tot}$ are shown in Figure S13. The calculated *ZT*s based on three measurements are shown in Figure S8, in which the three curves coincide with each other through a deviation of ~5%, confirming the thermal stability of the PM samples. In addition, three PM samples with a 90% relative density were prepared and tested in parallel, and the *T*-dependent *S*, $\sigma$, and $\kappa_{tot}$ are shown in Figure S14. The *ZT* curves for the three samples are shown in the inset of Figure S8, and a deviation of 7% between these samples proved the repeatable synthesis of the PM $Mg_3Sb_2$ based materials.

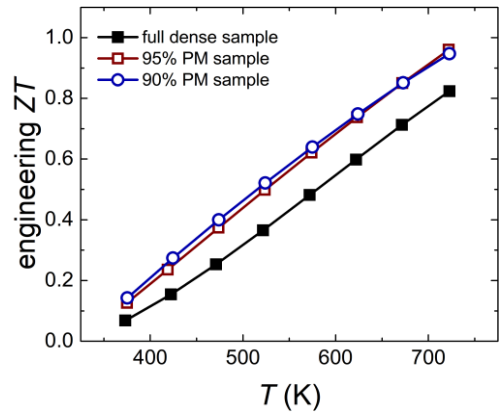

**Figure S9.**
Hot-side temperature-dependent engineering ($ZT$)eng of fully dense and PM samples (95% PM sample and 90% PM sample) (the cold-side temperature was set to 323 K).

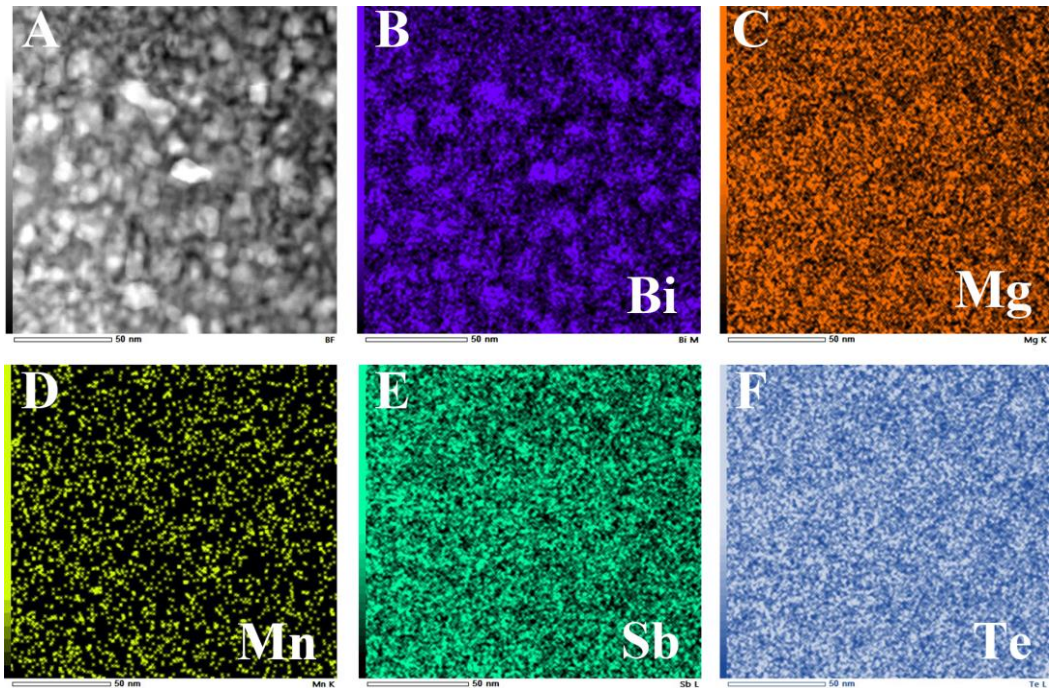

**Figure S10.**
(A) EDX mapping area. (B)–(F) the mapping results of compositional elements.

The Energy Dispersive X-ray Spectroscopy EDX element mapping was carried out, as shown in Figs. S11A through S11F, corresponding to Bi, Mg, Mn, Sb, and Te, respectively. The elemental mapping of Bi is found to fit well with the distribution of the precipitates in Figure S10A, whereas the mappings of Mg and Sb are in reverse coincidence to the precipitate distribution. Te was found to be uniformly distributed in the selected area, but Mn is accumulated somewhat in these areas probably because of its high melting temperature. EDX elemental mappings confirmed the coexistence of Bi nano-precipitates and the spinodal decomposition in the PM $Mg_{3.225}Mn_{0.025}Sb_{1.5}Bi_{0.49}Te_{0.01}$ sample, which is found to be favorable for the thermoelectric transport in this material.

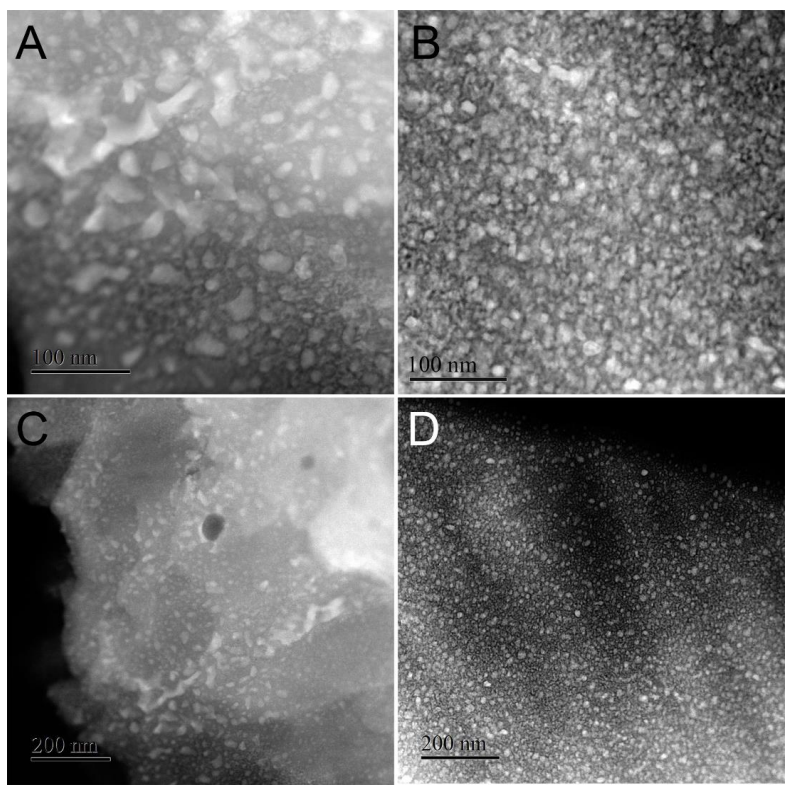

**Figure S11.**
Comparison of HRTEM images of the Bi nano-precipitates among (A, C) 95% PM sample and (B, D) 90% PM sample.

As shown in the Supporting Information (Figure S11), both the PM 95% and 90% samples exhibit a high density of precipitates, whereas the overall sizes of the precipitates in the PM 95% sample are larger than those in the 90% sample, revealing that the subtle difference in pressures applied on the PM $Mg_{3.225}Mn_{0.025}Sb_{1.5}Bi_{0.49}Te_{0.01}$ material during the SPS may play a critical role in the thermodynamic states and the microstructures of the final samples. It is known that $Mg_3Sb_2$ based materials become thermally unstable at up to 773 K because of the volatile constituent elements, and the solubility of Bi and Te in $Mg_3Sb_2$ is also limited by the equilibrium phase diagram, where $T$ is the dominant parameter. In our synthesis of the PM materials at a high temperature of 1073 K, the partial disassociation of Bi can occur primarily in the spinodal decomposition areas, which has already been formed during the low-temperature synthesis process in PM $Mg_{3.225}Mn_{0.025}Sb_{1.5}Bi_{0.49}Te_{0.01}$, as shown in the TEM images. These disassociated Bi species endotaxially grown in the matrix would significantly lower the mechanical strength of the $Mg_{3.225}Mn_{0.025}Sb_{1.5}Bi_{0.49}Te_{0.01}$ samples, leading to a softening of the disc at 1073 K. However the softened PM samples can be secured in the specially designed graphite die, accompanied by the porous structures during the cooling process, disassociated Bi species are transformed into endotaxially grown nano-precipitates in the matrix.

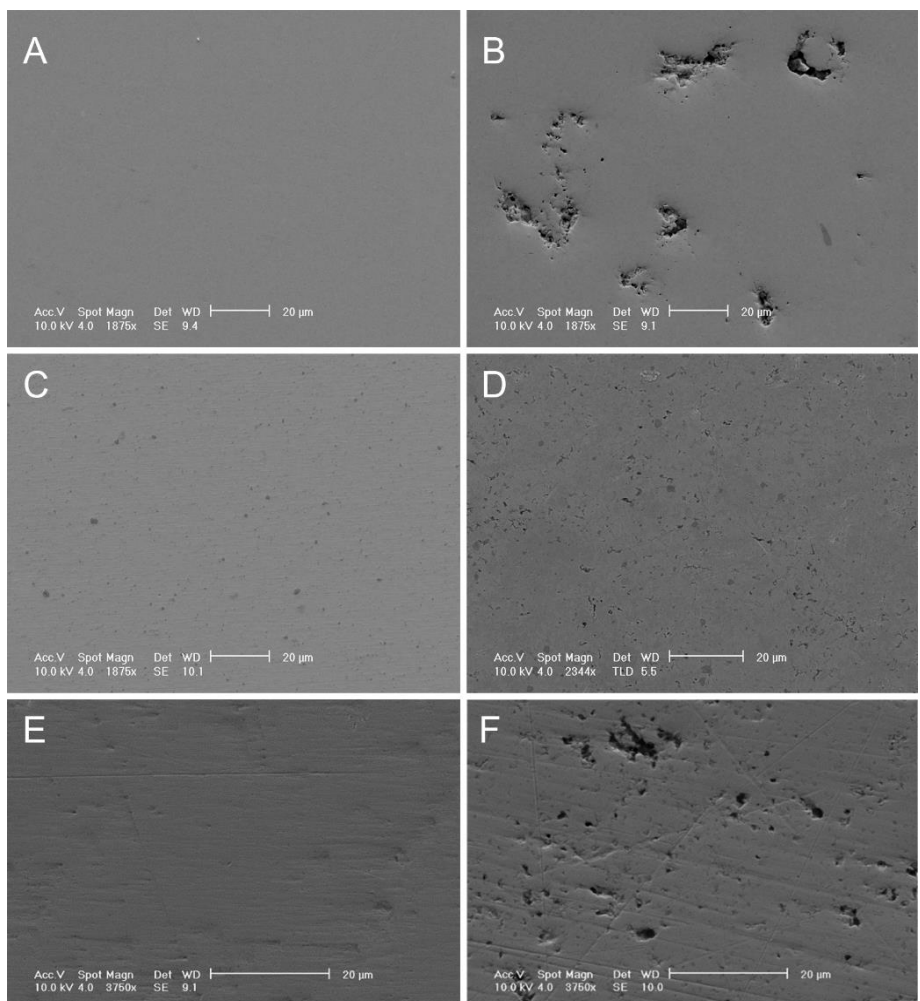

**Figure S12.**
SEM images of fully dense and PM samples. (A), (C), and (E) show fully dense samples of $Hf_{0.25}Zr_{0.75}NiSn_{0.99}Sb_{0.01}$, $Bi_{0.5}Sb_{1.5}Te_3$, and $FeNb_{0.56}V_{0.24}Ti_{0.2}Sb$, respectively. (B), (D), and (F) show the PM samples of $Hf_{0.25}Zr_{0.75}NiSn_{0.99}Sb_{0.01}$, $Bi_{0.5}Sb_{1.5}Te_3$, and $FeNb_{0.56}V_{0.24}Ti_{0.2}Sb$, respectively.

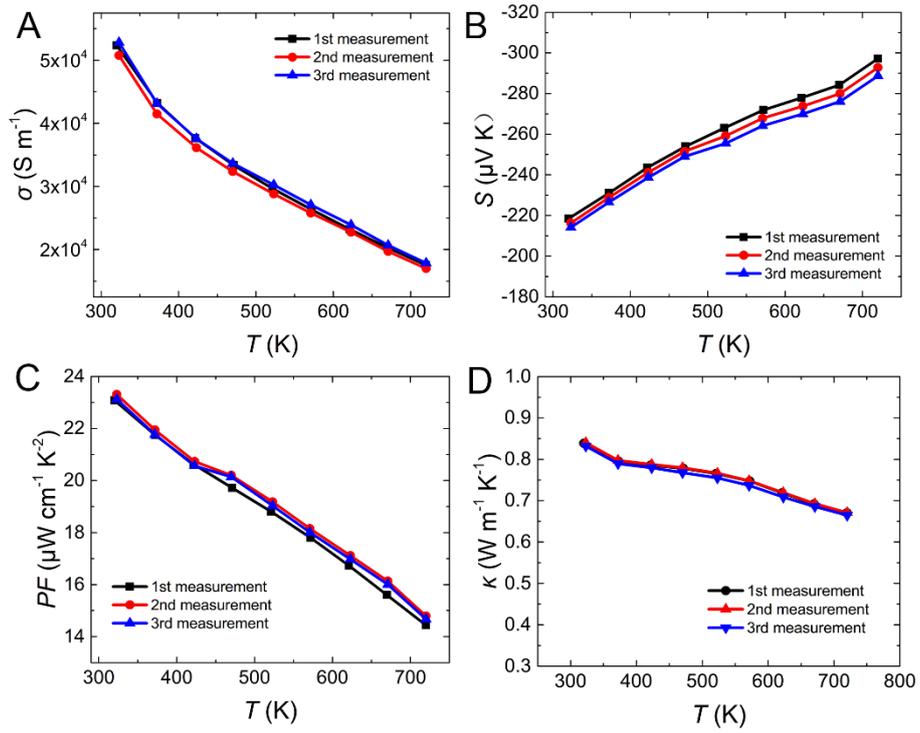

**Figure S13.**
Detailed thermoelectric properties based on three measurements for the same $Mg_{3.225}Mn_{0.025}Sb_{1.5}Bi_{0.49}Te_{0.01}$. (A) $T$ dependence of $\sigma$. (B) $T$ dependence of $S$. (C) $T$ dependence of power factor. (D) $T$ dependence of total thermal conductivity $\kappa$.

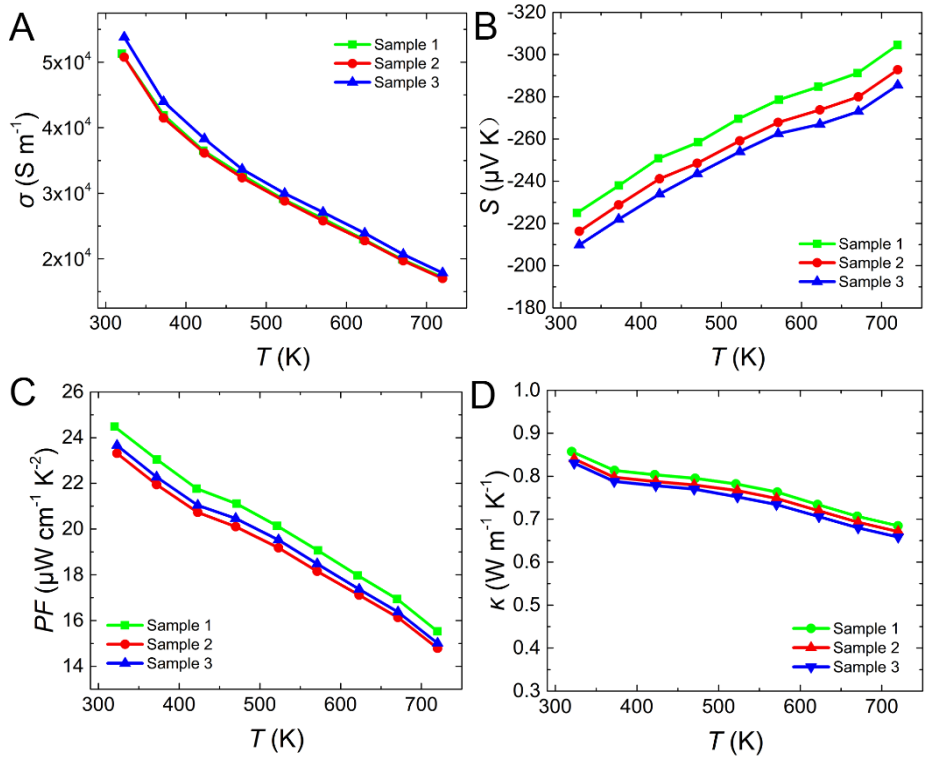

**Figure S14.**
Detailed thermoelectric properties for three different 90% PM samples of Mg$_{3.225}$Mn$_{0.025}$Sb$_{1.5}$Bi$_{0.49}$Te$_{0.01}$. (A) $T$ dependence of $\sigma$. (B) $T$ dependence of $S$. (C) $T$ dependence of power factor. (D) $T$ dependence of total thermal conductivity $\kappa$.

**Table S1.**

Actual compositions of fully dense sample (pressed at 873 K) and PM samples for nominal $Mg_{3.225}Mn_{0.025}Sb_{1.5}Bi_{0.49}Te_{0.01}$ material.

| Sample | Actual composition | | | | |
|---|---|---|---|---|---|
| | Mg | Mn | Sb | Bi | Te |
| Fully Dense sample | 3.164435 | 0.021521 | 1.483871 | 0.516129 | 0.011591 |
| 95% PM sample | 3.174603 | 0.022222 | 1.492063 | 0.507937 | 0.009524 |
| 90% PM sample | 3.223806 | 0.019355 | 1.516129 | 0.483871 | 0.009677 |

**Table S2.**

Calculated parameters for estimating the lattice thermal conductivity.

| Parameters | Fully Dense | 95% PM | 90% PM |
|---|---|---|---|
| Debye temperature, $\theta_D$(K) | 187 | 187 | 187 |
| Grüneisen parameter, $\gamma$ | 2.3526 | 2.0433 | 2.0 |
| Average atomic mass, M(g mol$^{-1}$) | 365.634 | 365.634 | 365.634 |
| Volume per atom, V(Å$^3$) | 130 | 130 | 130 |
| Average grain size, l(nm) | 370 | 3000 | 3000 |
| Disorder scattering parameter, $\Gamma$ | 8.21*10$^{-2}$ | 8.21*10$^{-2}$ | 8.21*10$^{-2}$ |
| Longitudinal phonon velocity, $v_l$(m s$^{-1}$) | 4709 | 3605 | 3476 |
| Transverse phonon velocity, $v_t$(m s$^{-1}$) | 2109 | 1779 | 1738 |
| nanoparticle average size, R（nm） | - | 1.7 | 2.2 |
| nanoparticle density, $d_p$(g/cm$^3$) | - | 5.18 | 5.18 |
| Density of state effective mass of conduction band, $m^*$($m_e$) | 0.19 | 0.19 | 0.19 |
| Deformation potential, $E_{def}$ (eV) | 8.49 | 4.24 | 4.42 |